\documentclass[twocolumn,showpacs,amsmath,amssymb,prl]{revtex4}

\usepackage{graphicx}
\usepackage{bm}

\begin{document}
\title{Non-adiabatic pumping through interacting quantum dots}

\author{Fabio Cavaliere$^1$, Michele Governale$^{2,3}$, and J\"urgen K\"onig$^2$}
 \affiliation{$^1$ LAMIA CNR-INFM, Dipartimento di Fisica, Universit\`a di Genova,  Via Dodecaneso 33, 16146 Genova, Italy\\
$^2$ Theoretische Physik, Universit\"at Duisburg-Essen and CeNIDE, 47048 Duisburg, Germany\\
$^3$ School of Chemical and Physical Sciences, Victoria University of 
Wellington, PO Box 600, Wellington 6140, New Zealand}

\date{\today}

\begin{abstract}
  We study non-adiabatic two-parameter charge and spin pumping through
  a single-level quantum dot with Coulomb interaction.  For the limit
  of weak tunnel coupling and in the regime of pumping frequencies up
  to the tunneling rates, $\Omega \lesssim \Gamma/\hbar$, we perform
  an exact resummation of contributions of all orders in the pumping
  frequency. As striking non-adiabatic signatures, we find
  frequency-dependent phase shifts in the charge and spin currents,
  which opens the possibility to control charge and spin currents by
  tuning the pumping frequency.  This includes the realization of an
  effective single-parameter pumping as well as pure spin without
  charge currents.
\end{abstract}

\maketitle

\noindent\textit{Introduction.} Pumping is a transport mechanism which
induces dc charge and spin currents in a nano-scale conductor in the
absence of a bias voltage by means of a time-dependent control of some
system parameters.  Recently there have been several experimental
works on pumping in nanostructures
\cite{pothier92,martinis,switkes99,watson,fletcher,fuhrer,buitelaar,kaestner}.
Theoretically, its interest lies in the possibility to investigate
non-equilibrium phenomena induced by the explicit time-dependence of a
nanoscale quantum system.  A lot of interest has been devoted to the
adiabatic regime, realized when the time-dependence of the parameters
is slow in comparison to the characteristic time scales of the system,
such as the dwell time. 
% MG commented out the following sentence 
%In the adiabatic limit, the pumped charge and
%spin do not depend on the detailed time evolution of the parameters
%but only on the area of the pumping cycle in parameter space
%\cite{brouwer98}. 
Many theoretical works have dealt with adiabatic pumping in systems
with weak electron-electron interaction
\cite{brouwer98,zhou99,buttiker01,makhlin01,buttiker02,entin02} as
well as in systems where the Coulomb interaction cannot be treated in
a mean-field approach
\cite{aleiner98,citro03,aono04,brouwer05,splett05,sela06,splett06,fioretto07,arrachea08,splett08}.
Pumping beyond the adiabatic limit, on the other hand, gives rise to
larger pumping currents, facilitating the experimental investigation
of this transport mechanism. Moreover, it adds another control
parameter, the pumping frequency $\Omega$, which can be used to steer
the charge and spin currents.  Pumping in the non-adiabatic regime is
intrinsically a strong-non-equilibrium phenomenon and its theoretical
description is challenging. In the limit of weak Coulomb interaction,
treated with a Hartree approach, a general theoretical framework can
be based on the Floquet scattering matrix
\cite{floquet}.  In systems with strong
Coulomb interaction, {\em no} such general framework exists. As a
paradigmatic system, we consider a single-level quantum dot. Pumping
in this type of systems has been studied either in the adiabatic
regime
\cite{aono04,splett05,sela06,splett06,fioretto07,arrachea08,splett08},
or in the opposite limit of very large frequency
\cite{hazelzet,cota05,sanchez,braun08}. Typically, the latter is
studied in the context of photon-assisted tunneling
\cite{bruder94,photon_exp}.  On the contrary, in the present Letter we
start from the low-frequency regime, including higher orders in the
pumping frequency employing a diagrammatic real-time transport
approach, which allows to include Coulomb interaction and
non-equilibrium effects. The transport quantities are then computed
perturbatively in the tunnel-coupling strengths. In the sequential
tunneling regime all orders in the pumping frequency $\Omega$ can be
resummed.\\
\noindent\textit{Model.}
We consider a single-level quantum dot, tunnel-coupled to two metallic
leads. The Hamiltonian of the system is
$H=H_{\text{dot}}+H_{L}+H_{R}+H_{\text{tunn}}$.  The dot is described
by the Anderson impurity model
\begin{equation} 
H_{\text{dot}}=\varepsilon(t)(n_{\uparrow}+n_{\downarrow})+U
n_{\uparrow}n_{\downarrow},  
\end{equation}
where $n_{\sigma}=d^{\dagger}_{\sigma}d_{\sigma}$ with the
annihilation operator $d_{\sigma}$ for an electron with spin
$\sigma=\uparrow,\downarrow$. The eigenstates of $H_{\text{dot}}$ are
$|\chi\rangle$ with $\chi\in\{0,\uparrow,\downarrow,d\}$,
corresponding to an empty, singly occupied with spin up or down, and
doubly occupied dot respectively. The level position $\varepsilon(t)$
is periodically modulated in time. Electrons in lead $\alpha=L,R$ are
described by $H_{\alpha}=\sum_{k,\sigma}\varepsilon_{\alpha
  k\sigma}c^{\dagger}_{\alpha k\sigma}c_{\alpha k\sigma}$. The right
lead can be ferromagnetic while the left one is non-magnetic. The spin
polarization is $p_R\equiv (\nu_{R+}- \nu_{R-})/( \nu_{R+}+
\nu_{R-})$, where $\nu_{R+}$ ($\nu_{R-}$) is the density of states at
the Fermi energy for the majority (minority) spins in the right lead
and $0\leq p_{R}\leq 1$. Dot and leads are coupled by
$H_{\text{tunn}}=\sum_{\alpha}V_{\alpha}(t)\sum_{k,\sigma}c^{\dagger}_{\alpha
  k\sigma}d_{\sigma}+\mathrm{H.c.}$; the tunnel amplitudes
$V_{\alpha}(t)$ vary in time. The tunnelling strength is
$\Gamma_{\alpha}(t)=\sum_{\sigma}\Gamma_{\alpha\sigma}(t)/2$ with
$\Gamma_{\alpha\sigma}(t)=2\pi\nu_{\alpha\sigma}|V_{\alpha}(t)|^{2}$. The
total tunnelling strength is
$\Gamma(t)=\Gamma_{L}(t)+\Gamma_{R}(t)$. No voltage is applied: charge
and spin currents arise only due to the periodic modulations of
$\varepsilon(t)$ and $\Gamma_{\alpha}(t)$, denoted collectively by
$X(t)$ in the following, with frequency
$\Omega$.\\
\noindent\textit{Non--adiabatic pumping.} The dot is described by its
reduced density matrix $\rho_{\text{dot}}(t)$.  In the present case,
the dynamics of the diagonal and off-diagonal elements of
$\rho_{\text{dot}}(t)$ are decoupled, i.e., we can restrict
ourselves to study the occupation probabilities,
$P_{\chi}(t)=\langle\chi|\rho_{\text{dot}}|\chi\rangle$, whose
time evolution is governed by a generalized master equation
\begin{equation}
\label{eq:gme}
\dot{\mathbf{P}}(t)=\int_{-\infty}^{t}dt'
\mathbf{W}(t,t')\cdot\mathbf{P}(t'), 
\end{equation} where
$\mathbf{P}=(P_{0},P_{\uparrow},P_{\downarrow},P_{d})^{T}$. 
The kernel $\mathbf{W}(t,t')$ functionally depends on
$X(t)$. We perform a series expansion in powers of the pumping frequency.
Generalizing the adiabatic expansion of Ref.~\cite{splett06}, we keep {\em all} orders in $\Omega$. For this, we write $\mathbf{P}(t)=\sum_{k\geq 0}\mathbf{P}^{(k)}_t$ and $\mathbf{W}(t,t')=\sum_{k\geq 0}\mathbf{W}_{t}^{(k)}(t-t')$, where the superscript $(k)$ indicates terms of order $\Omega^k$. In $\mathbf{W}_{t}^{(k)}(t-t')$, the time dependence of all parameters is expanded to order $k$ around the final time $t$ and only terms of order $k$ in the time derivatives are retained. Expanding $\mathbf{P}(t')$ in Taylor series and performing a Laplace transform of the r.h.s. of Eq.~(\ref{eq:gme}), we obtain
\begin{equation}
\label{eq:gme2}
\hbar\sum_{n=0}^{\infty}\frac{d\mathbf{P}_{t}^{(n)}}{dt}=\sum_{p,q,k=0}^{\infty}\frac{1}{k!}\left(\partial^{k}\mathbf{W}_{t}^{(p)}\right)\cdot\frac{d^{k}\mathbf{P}_{t}^{(q)}}{dt^{k}}\, ,
m\end{equation}
where $\partial^{k}{\mathbf{W}}_{t}^{(p)}=\lim_{z\to
  0^{+}}\partial^{k}{\mathbf{W}}_{t}^{(p)}(z)/\partial z^{k}$ and $\mathbf{W}_{t}^{(p)}(z)=\hbar\int_{-\infty}^t e^{-z
  (t-t')}\mathbf{W}_{t}^{(p)}(t-t')dt'$ is the Laplace transformed
kernel.
How to count the time derivatives of $\mathbf{P}$ in this expansion
depends on the considered frequency regime. In this paper, we consider
the regime $\hbar\Omega\lesssim\Gamma$, for which the system quickly
relaxes to an oscillatory steady state with the frequency of $X(t)$,
i.e., each time derivative introduces one power in $\Omega$. 
% MG commented out the following sentence
% In the opposite regime, $\hbar\Omega\gtrsim\Gamma$, the most relevant
% variation of $\mathbf{P}(t)$ would occur on the relaxation time scale
% $\hbar \Gamma^{-1}$, and no extra powers in $\Omega$ would appear.

In addition to the expansion in frequency, we perform a systematic
expansion of $\mathbf{W}_{t}^{(k)}=\sum_{j\geq
  1}\mathbf{W}_{t,j}^{(k)}$ and $\mathbf{P}_{t}^{(k)}=\sum_{j\geq
  -k}\mathbf{P}_{t,j}^{(k)}$ in powers of the tunnel-coupling strength
$\Gamma$. The order in $\Gamma$ is indicated by the subscript $j$.
Upon substitution into Eq.~(\ref{eq:gme2}), the orders of $\Omega$ and
$\Gamma$ on both sides are matched, giving rise to a {\em hierarchy}
of coupled equations for $\mathbf{P}_{t,j}^{(k)}$.  This matching
requires the expansion for $\mathbf{P}_{t}^{(k)}$ to start from
$\Gamma^{-k}$. In the rest of the paper, we concentrate on the limit
of weak tunnel coupling, i.e., we compute the kernel to first order in
$\Gamma$.  In this case, the hierarchy of equations reduces to
\begin{equation}
0=\mathbf{W}_{t,1}^{(0)}\cdot\mathbf{P}_{t,0}^{(0)}\ ;\ \hbar\dot{\mathbf{P}}_{t,-k}^{(k)}=\mathbf{W}_{t,1}^{(0)}\cdot\mathbf{P}_{t,-(k+1)}^{(k+1)} \, .
\label{eq:hie}
\end{equation}
for $k\geq 0$.  Remarkably, only the instantaneous kernel
$\mathbf{W}_{t,1}^{(0)}$, corresponding to freezing the time evolution
of $X(t)$ at time $t$, is needed. The rules for evaluating this kernel
are given in Ref.~\cite{splett08}. We can solve for
$\mathbf{P}_{t,-k}^{(k)}$ recursively starting from the instantaneous
term $\mathbf{P}_{t,0}^{(0)}$.  The charge and and spin currents in
the the left lead can be expanded in powers of $\Omega$ as well,
$I_{\xi}(t)=\sum_{k\geq 0}I_{\xi}^{(k)}(t)$ ($\xi=Q,S$).  The $k$-th
contribution is given by
$I_{\xi}^{(k)}(t)=(c_{\xi}/\hbar)\mathbf{e}^{T}\cdot\mathbf{W}_{t,1}^{\xi,L(0)}\cdot\mathbf{P}_{t,-k}^{(k)}$,
with $c_{Q}=e$, $c_{S}=\hbar/2$, and
$\mathbf{e}^{T}=(1,\ldots,1)$. The symbols
$\mathbf{W}_{t,1}^{\xi,L(0)}$ stand for the current rates, which take
into account the number of electrons transferred to the left
lead~\cite{splett08}. In steady state, the average pumped charge and
spin currents per period are $\mathcal{I}_{\xi}=\frac{\Omega}{2
  \pi}\int_{0}^{2\pi/\Omega}dt I_{\xi}(t)$. It is immaterial which
barrier is chosen for calculating the average pumped currents per
period, since the current continuity equation is fulfilled for each
order of the expansions.

The procedure outlined above is very general, provided the
weak-coupling limit. To proceed in our specific model, we introduce
the dot charge (in units of $e$) $N_Q (t)=(0,1,1,2)\cdot\mathbf{P}(t)$
and spin (in units of $\hbar/2$) $N_S
(t)=(0,1,-1,0)\cdot\mathbf{P}(t)$ and their deviations from the
instantaneous values $\Delta N_{\xi}=N_{\xi} (t)-N_{\xi}^{(0)}(t)$
with $N_{Q}^{(0)}(t)=2f[\varepsilon(t)]/
\{1+f[\varepsilon(t)]-f[\varepsilon(t)+U]\}$ and $N_{S}^{(0)}(t)=0$. A
resummation of Eqs.~(\ref{eq:hie}) yields
\begin{equation}
\Delta\dot{ N}_{\xi}+\tau_{\xi}^{-1}(t)\Delta N_{\xi}+p_R\tau_{\xi}^{-1}(t)\gamma_{R}(t)\Delta N_{\bar{\xi}}=-\dot{N}_{\xi}^{(0)}(t)\label{eq:eomQ} ,
\end{equation}
where $\gamma_{\alpha}(t)=\Gamma_{\alpha}(t)/\Gamma(t)$ and $\bar{\xi}= S$ if $\xi=Q$ and vice versa. 
The two time scales $\tau_\xi(t)$, defined by 
\begin{eqnarray}
\hbar\tau_{Q}^{-1}(t)&=&\Gamma(t)\left\{1+f[\varepsilon(t)]-f[\varepsilon(t)+U]\right\}
\\
\hbar\tau_{S}^{-1}(t)&=&\Gamma(t)\left\{1-f[\varepsilon(t)]+f[\varepsilon(t)+U]\right\} \, ,
\end{eqnarray}
mare the instantaneous charge and spin relaxation times for $p_R=0$,
with $f(E)$ being the Fermi function. Finally, the charge and spin
currents can be recast in the form
\begin{equation}
I_{\xi}(t)=-c_{\xi}\tau_{\xi}^{-1}(t)\gamma_{L}(t)\Delta N_{\xi}(t)\label{eq:currents}\, .
\end{equation}

Equations~(\ref{eq:eomQ}) and (\ref{eq:currents}) constitute the main
result of this Letter. They allow to evaluate the non--adiabatic
pumped charge and spin currents for frequencies $\hbar\Omega \lesssim
\Gamma$. Solutions for both $p_R =0$ and $p_R\neq 0$ are discussed in
the following. As a specific pumping model, we choose
$\Gamma_{L}(t)=\bar{\Gamma}_{L}+\Delta\Gamma_{L}\sin{(\Omega t)}$ and
$\varepsilon(t)=\bar{\varepsilon}+\Delta\varepsilon\sin{(\Omega
  t+\phi)}$, with $\phi$ the pumping phase.  We take $\Gamma_{R}$ as
time independent. 
% FC streamlined a little bit here
We focus chiefly on the case of weak pumping
$\Delta\Gamma_{L},\Delta\varepsilon\ll\Gamma$ which allows a
analytical treatment. However, we want to stress that
Eqs.~(\ref{eq:eomQ}--\ref{eq:currents}) are not restricted to this
regime. Numerical results for strong pumping will be discussed in the
last part of this work.

In the weak-pumping regime, the charge and spin currents have the form
\begin{equation}
\mathcal{I}_{\xi}=\mathcal{I}_{\xi}^{\text{max}}\sin{(\phi+\Delta\phi_{\xi})} \, ,
\end{equation} 
where both the amplitude $\mathcal{I}_{\xi}^{\text{max}}$ and the
phase shift $\Delta\phi_{\xi}$ are odd functions of $\Omega$, i.e.,
when expanding the currents in powers of the pumping frequency, all
the odd powers of the frequency $\Omega$ are proportional to
$\sin(\phi)$ while the even powers are proportional to $\cos(\phi)$.
The analytical expressions for the $\Omega$-dependent amplitudes are,
for arbitrary value of the spin polarization $p_R$, rather lengthy and
we do not report them here.

The zeroth order in $\Omega$ describes the instantaneous equilibrium,
for which both the average charge and spin current vanish.  Adiabatic
pumping corresponds to expanding the currents to first order in
$\Omega$, which leads to $\mathcal{I}_{\xi}^{\rm ad}=(\partial
\mathcal{I}_{\xi}^{\text{max}}/ \partial \Omega )|_{\Omega=0}
\Omega\sin \phi$.  The non-adiabatic contributions to the pumped
charge and spin do not only introduce an $\Omega$-dependence of the
amplitude, but change the pumping behavior qualitatively.  First,
phase shifts $\Delta\phi_{\xi}$ are generated, i.e., pumping is also
possible for $\phi=0$ or $\pi$, which corresponds to {\em
  single-parameter pumping}.  Second, the phase shifts for charge and
spin pumping may differ from each other.  As we will see below, this
allows pure spin pumping~\cite{mucciolo} without charge pumping. 
 % FC Added the following sentence. 
Phase shifts are a general feature of non-adiabatic, driven quantum
dynamics and are found in disparate contexts, from
circuit-QED~\cite{walraff} to driven optical lattices~\cite{gommers}
and stochastic quantum resonance~\cite{grifoni}. Molecular
systems~\cite{astumian} and nano-electomechanical
systems~\cite{pistolesi} are especially interesting in connection to our technique.\\
%Here is MG's original one :)
%Phase shift are a general feature of non-adiabatic, driven quantum
%dynamics and the method introduced here can be employed to investigate
%them also in different context, such as
%molecular systems \cite{astumian}, nano-electomechanical systems \cite{pistolesi}, and circuit-QED~\cite{walraff}.\\
%%
\noindent\textit{Non-magnetic case.} 
In the limit of non-magnetic leads, $p_R=0$, the expressions for the
charge and spin currents simplify substantially.  Then, the spin
currents vanishes, $\mathcal{I}_{S}\equiv 0$, and for the charge
current we find
\begin{eqnarray}
(2e\bar{\Gamma}_{L})\mathcal{I}_{Q}^{\text{max}}&=&\Delta\Gamma_{L}\Delta\varepsilon G^{\rm lin} \Omega\bar{\tau}_{Q}[1+(\Omega\bar{\tau}_{Q})^2]^{-\frac{1}{2}}\label{eq:amplitude}
\\
\Delta\phi_Q&=&-\arctan{(\Omega\bar{\tau}_{Q})}\label{eq:shift} \, .
\end{eqnarray}
Here, we used the expression for the linear DC conductance $G^{\rm
  lin}=\frac{2e^2}{\hbar}\beta\frac{\bar\Gamma_L\bar\Gamma_R}{\bar\Gamma}
\frac{f(\bar{\varepsilon})[1-f(\bar{\varepsilon}+U)][1-f(\bar{\varepsilon})+f(\bar{\varepsilon}+U)]}{1+f(\bar{\varepsilon})-f(\bar{\varepsilon}+U)}$,
which, for $k_{B}T \ll U$ has two maxima at
$\bar{\varepsilon}=k_{B}T\ln{2}$ and
$\bar{\varepsilon}=-U-k_{B}T\ln{2}$.

\begin{figure}[htpb]
\begin{center}
\includegraphics[height=8cm,keepaspectratio]{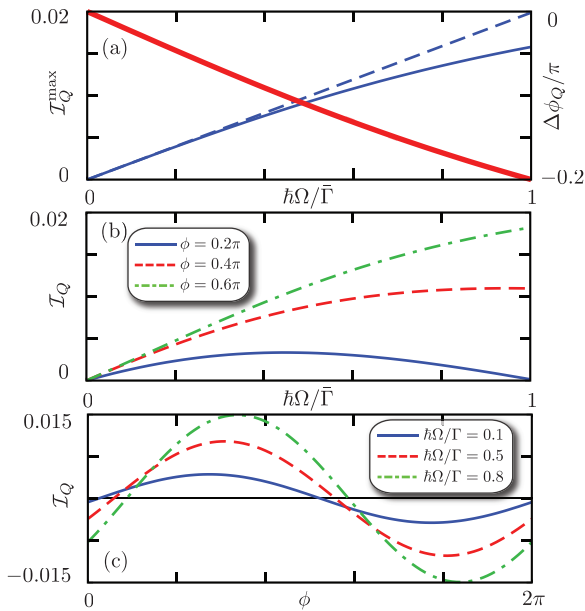}
\caption{(Color online) (a) Charge-current amplitude
  $\mathcal{I}_{Q}^{\text{max}}$ (thin solid), its adiabatic
  approximation $\mathcal{I}_{Q}^{\text{max, ad}}$ (thin dashed), and
  phase shift $\Delta\phi_Q$ (thick) as a function of
  $\hbar\Omega/\bar{\Gamma}$. (b) Pumped current $\mathcal{I}_{Q}$ as
  a function of $\hbar\Omega/\bar{\Gamma}$, for $\phi=0.2\pi$ (solid),
  $\phi=0.4\pi$ (dashed), $\phi=0.6\pi$ (dash-dotted).  (c) Pumped
  charge current $\mathcal{I}_{Q}$ as a function of $\phi$, for
  $\hbar\Omega/\Gamma=0.1$ (solid), $0.5$ (dashed) and $0.8$
  (dash-dotted).  All currents in units of
  $e\beta\Delta\Gamma_{L}\Delta\varepsilon/\hbar$.  Other parameters:
  $\bar{\varepsilon}=k_B T \ln{2}$, $U=9 \bar{\Gamma}$,
  $k_{B}T=3\bar{\Gamma}$, $p_R=0$, and
  $\bar{\gamma}_R=1/2$.}\label{fig:fig1}
\end{center}
\end{figure}
In the following, we concentrate on the peak at
$\bar{\varepsilon}=k_{B}T\ln{2}$. In Fig.~\ref{fig:fig1}(a) the
charge-current amplitude $\mathcal{I}_{Q}^{\text{max}}$ and the phase
shift $\Delta\phi_{Q}$ are shown (solid lines) as a function of
$\hbar\Omega/\bar{\Gamma}$.  The amplitude is reduced as compared to
the adiabatic approximation (dashed line).  In Fig.~\ref{fig:fig1}(b)
the pumped current $\mathcal{I}_{Q}$ is plotted as a function of the
frequency $\Omega$ for different values of the pumping phase $\phi$.
Deviations from the adiabatic approximation, simply given by the
tangents at $\Omega=0$, become most pronounced as the pumping phase
$\phi$ approaches zero: in this limit the phase shift
$\Delta\phi_{Q}$, a striking signature of non-adiabatic pumping, is
most important. Figure~\ref{fig:fig1}(c) shows $\mathcal{I}_{Q}$ as a
function of the phase $\phi$ for different values of
$\hbar\Omega/\bar{\Gamma}$.  The growing relevance of the phase shift
$\Delta \phi_Q$ as the pumping frequency is increased is apparent. As
commented above, its most important consequence is the possibility to
obtain an effective one-parameter pumping for $\phi=0,\pi$.

The frequency dependence of the phase shift $\Delta\phi_{Q}$ is that
of a low--pass filter with cutoff frequency $\bar{\tau}_{Q}^{-1}$. It
signals the reduced ability of fast pumps to transfer charge due to
the non--zero response time of the dot to charge excitations,
$\bar{\tau}_{Q}$.
  
\begin{figure}[htbp]
\begin{center}
\includegraphics[height=8cm,keepaspectratio]{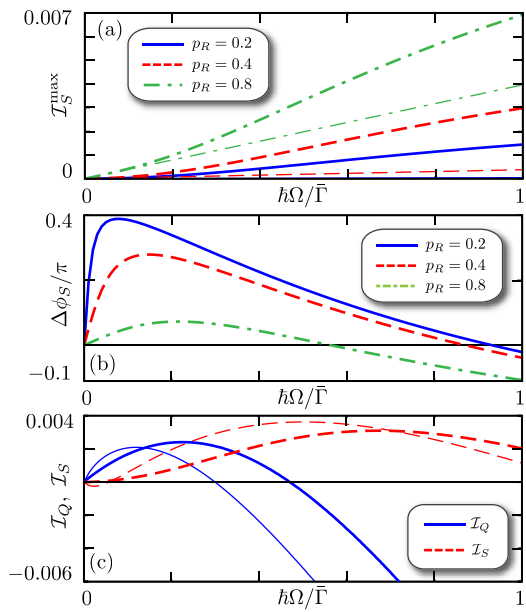}
\caption{(Color online) (a) Spin-current amplitude
  $\mathcal{I}_{S}^{\text{max}}$ (thick) and its adiabatic
  approximation $\mathcal{I}_{S}^{\text{max,ad}}$ (thin) as a function
  of $\hbar\Omega/\bar{\Gamma}$ for different values of $p_R$. (b)
  Phase shift $\Delta\phi_{S}$ as a function of
  $\hbar\Omega/\bar{\Gamma}$ for different values of $p_R$.(c) Charge
  and spin currents $\mathcal{I}_{Q,S}$ as a function of
  $\hbar\Omega/\bar{\Gamma}$ with $p_R=0.4$ and $\phi=\pi/9$ for weak
  pumping (thick) and strong pumping (thin, with
  $\Delta\Gamma_{L}=\Delta\varepsilon=0.4\bar{\Gamma}$). Spin currents
  in units of $\beta\Delta\Gamma_{L}\Delta\varepsilon/2$. Other
  parameters: $\bar{\varepsilon}=k_B T\log{2}$, $U=9 \bar{\Gamma}$,
  $k_B T=3 \bar{\Gamma}$, and $\bar{\gamma}_R=1/2$.}\label{fig:fig2}
\end{center}
\end{figure}

\noindent\textit{Ferromagnetic case.} 
% MG streamlined the following text to gain space
%The lengthy analytical expressions for the charge and spin currents in
%the weak-pumping regime for $p_R\neq 0$ will not be reported here.
%Instead, we resort to a graphical presentation of the average pumped
%currents. 
%
In the weak-pumping regime for $p_R\neq 0$, we resort to a graphical presentation of the average pumped currents.  In Fig.~\ref{fig:fig2}(a) and (b) the spin-current
amplitude $\mathcal{I}_{S}^{\text{max}}$ and phase shift
$\Delta\phi_{S}$ is shown as a function of the frequency $\Omega$.
Both exhibit a strong dependence on the polarization $p_R$ of the
ferromagnet.
% MG replaced "On the other hand" with "By contrast"
By contrast,  
% On the other hand, 
the charge amplitude and phase shift
(both not shown here) are almost insensitive to $p_R\neq0$. In
particular, $\Delta\phi_{Q}$ is essentially given by
Eq.~\ref{eq:shift} and charge transfer is still dominated by the
cutoff frequency $\bar{\tau}_{Q}^{-1}$. 
On the other hand, the
intricate, coupled dynamics of spin and charge results in the spin
phase being given by
$\Delta\phi_{S}=\arctan(\Omega/2\Omega_{0})-\sum_{q=\pm}\arctan(\Omega/2\Omega_{q})$.
For $\bar{\gamma}_{R}=1/2$, the cut-off frequencies are
$\Omega_{0}=p_{R}^{2}\bar{\tau}_{S}^{-1}$ and
$\Omega_{\pm}=\tau_{Q}^{-1}+\tau_{S}^{-1}\pm\sqrt{(\tau_{Q}^{-1}-\tau_{S}^{-1})^{2}+p_{R}^{2}(\bar{\tau}_{Q}\bar{\tau}_{S})^{-1}}$. The
competition between $\Omega_{0}$ and the ``hybridized'' $\Omega_{\pm}$
gives rise to the non--monotonic behavior as well as the sign change
of $\Delta\phi_{S}$. The most striking consequence of
non--adiabaticity in the ferromagnetic case is the possibility to
obtain a {\em pure spin current}, $\mathcal{I}_{S}\neq0$ with
$\mathcal{I}_{Q}=0$, by tuning $\Omega$ such that
$\phi+\Delta\phi_{Q}(\Omega)=0$. This behavior is illustrated in
Fig.~\ref{fig:fig2}(c), thick lines. The adiabatic approximation,
missing the phase shifts, is unable to capture this effect. Pure spin
pumping, as all the effects induced by the ferromagnetic lead, are enhanced for $\bar{\gamma}_{R}\to 1$.\\
We conclude discussing the experimental observability of the effects
discussed above. Currents in the weak pumping regime are generally
small. However, our theory is valid also for strong pumping where
currents are larger and we have checked that {\em all} the results are
not qualitatively modified. As an example, Fig.~\ref{fig:fig2}(c)
(thin lines) shows the charge and spin currents for
$\Delta\bar{\Gamma}_{L}=\Delta\varepsilon=0.4\bar{\Gamma}$.  For a
temperature $T$ of the order of 1 K ($k_{B}T\approx 90\mu$ eV) and $
\bar{\Gamma}=k_B T/3$ , one has
$e\beta\Delta\Gamma_{L}\Delta\varepsilon/\hbar\approx 500$ pA as the
current unit. Clearly, pure spin pumping is still present.  The charge
current, ranging between -3 pA and 1 pA, is easily measurable. 
%% MG Added the following sentence
However, care needs to be exerted to minimize rectification effects due to stray capacitances 
and to distinguish them from pumping.\\
\textit{Conclusions.}
We have studied non-adiabatic two-parameter charge and spin pumping
through an interacting quantum dot, tunnel coupled to metallic
leads. In the sequential-tunneling regime, for frequencies smaller
than the tunnel rates, all orders in the pumping frequency can be
resummed exactly.  We find frequency-dependent phase shifts for the
pumped charge and spin current, neglected in the adiabatic
approximation.  They allow an effective single-parameter pumping,
i.e., finite pumping currents when the two pumping parameters
oscillate in phase or anti-phase. When one lead is ferromagnetic, the
pumping frequency turns out to be an important parameter to control
the spin current. In particular, due to the non-adiabaticity of
pumping, pure spin currents can be generated. The effects discussed
above can be experimentally verified in the strong-pumping regime,
where the pumped currents are sizable.\\
\textit{Acknowledgements.} We acknowledge financial support from the
DFG via SPP 1285 and via SFB 491, and from INFM-CNR via Seed Project
PLASE001.

\end{document}